# Nernst effect in the electron-doped cuprate superconductor La$_{2-x}$Ce$_x$CuO$_4$

P. R. Mandal[1, 2*], Tarapada Sarkar[1, 2*], J. S. Higgins[1, 2], Richard L. Greene[1, 2]

[1]*Center for Nanophysics & Advanced Materials, University of Maryland, College Park, Maryland 20742, USA.*
[2]*Department of Physics, University of Maryland, College Park, Maryland 20742, USA.*

We report a systematic study of the Nernst effect in films of the electron doped cuprate superconductor La$_{2-x}$Ce$_x$CuO$_4$ (LCCO) as a function of temperature and magnetic field (up to 14 T) over a range of doping from underdoped (*x*=0.08) to overdoped (*x*=0.16). We have determined the characteristic field scale H$_{C2}$* of superconducting fluctuation which is found to track the domelike dependence of superconductivity (T$_C$). The fall of H$_{C2}$* and T$_C$ with underdoping is most likely due to the onset of long range antiferromagnetic order. We also report the temperature onset, T$_{onset}$, of superconducting fluctuations above T$_C$. For optimally doped *x*=0.11 T$_{onset}$ ($\cong$ 39 K) is high compared to T$_C$ (26 K). For higher doping T$_{onset}$ decreases and tends to zero along with the critical temperature at the end of the superconducting dome. The superconducting gap closely tracks H$_{C2}$* measured from the temperature and field dependent Nernst signal.

Correspondence: rickg@umd.edu

*These authors have contributed equally to this work.

I. **Introduction**

The nature of the normal state and the origin of the high $T_C$- superconductor (HTSC) in the cuprates is still a major unsolved problem. In the hole-doped cuprates a mysterious "pseudo gap" is found, whose origin and relation to the HTSC is still not understood. Some early Nernst effect experiments [1] on underdoped p-type cuprates suggested that there is a significant temperature range above $T_C$ where superconductivity (SC) fluctuations exist. This was supported by theory [2] and other experiments [3, 4]. But, recently, the early interpretation of the Nernst effect has been questioned [5] and a much smaller temperature range of fluctuations has been proposed. Therefore, the issue of the range of SC fluctuations and their relation to the onset of the pseudogap in p-type cuprates is still controversial.

The electron-doped cuprates have a much simpler phase diagram with no p-type pseudogap. Prior Nernst effect experiments on n-type cuprates are basically in agreement that SC fluctuations occur over a rather narrow T range above $T_C$ at all doping [1]. Surprisingly, a rather large normal-state (quasiparticle) Nernst effect was observed [6] in the overdoped regime where one might expect a small effect because only hole carriers dominate the transport properties there (i.e. large hole-like Fermi surface). Stated another way, the Fermi Surface reconstruction (FSR) in the n-type cuprates does not appear to impact the normal-state Nernst effect in any dramatic way, whereas the Hall effect and thermopower show large changes at the FSR doping [6, 7].

In this paper we present Nernst effect experiments on another n-type cuprate, $La_{2-x}Ce_xCuO_4$ (LCCO). This research is motivated by the fact that LCCO can be doped over a wider range than other n-type cuprates. This allows us to examine the Nernst effect at doping throughout the SC dome including the FSR doping and to determine the temperature range of SC fluctuations over the entire SC dome.

The generation of a transverse electric field by a longitudinal thermal gradient and a perpendicular magnetic field, the Nernst effect, has attracted considerable attention in the cuprates due to observation of a large Nernst signal in the normal state. The Nernst effect is large in the superconducting state due to vortex motion but the normal state quasiparticle contribution is usually small (except in two band materials). The surprisingly large Nernst signal well above $T_C$ in the underdoped p-type cuprates was attributed to SC phase fluctuations and vortexlike excitations [8, 1]. The basic assumption of this interpretation was that, the usual SC Gaussian amplitude fluctuations have a weak contribution to the Nernst signal above $T_C$. But, in 2006 Pourret et al. showed the existence of fluctuating Cooper pairs up to a high temperature (T≈ 30×$T_C$) and higher magnetic field (H≈4×$H_{C2}$) for a conventional superconductor $Nb_xSi_{1-x}$ [9, 10]. For $Nb_xSi_{1-x}$ and another amorphous superconductor, $InO_x$, it was found that the fluctuations were in quantitative agreement with Gaussian fluctuations theory for a two-dimensional (2D) dirty superconductor [2]. The theory was restricted to low magnetic fields in the vicinity of the critical temperature. Later Michaeli and Finkel'stein developed theoretical calculations which explained quantitatively the experimental data of $Nb_xSi_{1-x}$ at temperature much higher than the critical temperature and at magnetic field much higher than the upper critical field [11]. Still later, Taillefer and his collaborators [5, 12] studied the Nernst effect in the hole-doped cuprate $La_{1.8-x}Eu_{0.2}Sr_xCuO_4$ and the electron-doped cuprate $Pr_{2-x}Ce_xCuO_4$ and concluded that quasiparticles and Gaussian fluctuations could explain the Nernst signal above $T_C$ in the under-doped as well as the over-doped region. Moreover, the temperature range of these fluctuations in the normal state was much smaller than suggested by the work of Wang et al [1].

There have been several prior Nernst effect experiments in different electron doped cuprates. In 2003 Balci et al. [13] observed a large normal state Nernst signal in $Pr_{2-x}Ce_xCuO_4$ (PCCO) films with Ce concentration varied around the optimal doping, which was in agreement with previous results in $Nd_{2-x}Ce_xCuO_4$ (NCCO) where the results were interpreted as the evidence for the existence of two carrier transport [14]. In 2007 Li et al. reported a systematic study of the normal state Nernst effect in PCCO over a wide range of Ce concentration to investigate the transport properties in the very underdoped and overdoped region [6]. The surprisingly large Nernst signal observed in overdoped and lightly underdoped PCCO is of uncertain origin but could be explained a two-band model. Recently Tafti et al. measured the Nernst signal in PCCO at four concentrations, from underdoped ($x = 0.13$) to overdoped ($x = 0.17$), for a wide range of temperatures above the critical temperature and showed that the data are quantitatively consistent with the theory of Gaussian superconducting fluctuations [12].

The detailed studies of the Nernst effect in electron-doped cuprates over a wide range of Ce concentration are important in order to understand the transport properties in the very underdoped or overdoped regimes. In this paper we study the vortex Nernst effect and superconducting fluctuations in the films of the electron-doped cuprate superconductor $La_{2-x}Ce_xCuO_4$ (LCCO) for both sides of the dome. The major advantage of using LCCO films is that we can study the Nernst effect up to the Fermi liquid regime i.e. the more overdoped side ($x = 0.21$). Here we have studied that how the SC fluctuations vary from underdoped to overdoped regime. Since optimal doping (0.11) [15] and the doping where the Fermi surface reconstruction occurs (0.14) are different, we can study the impact of the Fermi surface reconstruction on the Nernst effect. Another advantage is that a very low magnetic field of 14 T is enough to supress the superconducting fluctuations for LCCO films. We have found that the Nernst signal follows the Gaussian fluctuation theory for underdoped, optimal doped and over-doped samples above the critical temperature, which is consistent with the previous results found in PCCO [12]. Also from our measurements we have determined Ce concentration dependence of the upper critical field $H_{C2}*$, which characterizes the strength of the superconductivity. We found that $H_{C2}^*$ exactly tracks the $T_C$ dome. We also determined the extent of the SC fluctuation region above $T_C$ for all doping.

## II. Experimental details

The measurements have been performed on $La_{2-x}Ce_xCuO_4$ thin films for underdoped ($x$=0.08), optimally doped ($x$=0.11), and overdoped ($x$=0.13, 0.14, 0.15, and 0.16) compositions. The thin films were fabricated on (100) $SrTiO_3$ (10×5 mm$^2$) substrates by a Pulsed Laser deposition technique utilizing a KrF excimer laser [6]. The thickness of the samples used for this study is between 150 to 200 nm. The Nernst effect has been measured using a one heater-two thermometer technique. One end of the sample is attached to a copper block with a mechanical clamp and other end is left free. On the free end a small chip resistor heater is attached and two tiny Lakeshore Cernox thermometers are on the two ends of the sample to monitor the temperature gradient (0.7-1 K) continuously. The Nernst voltage measurements have been performed using a Keithley 2001 multimeter with sensitivity of several nanovolts by applying a magnetic field between 14 T and -14 T in a Physical Property Measurement System (PPMS). The measurements were done under high vacuum and the magnetic field has been applied perpendicular to *ab* plane. The sample temperature is taken as the average of hot and cold end temperatures. The final Nernst signal is obtained by subtracting the Nernst data at negative field from the Nernst data at positive field to eliminate any possible contribution from the Seebeck effect.

## III. Results

### 1. *In plane resistivity*

A standard four probe method has been used to measure the resistivity ρ of LCCO thin films at zero magnetic field. To understand the superconducting phase diagram we have studied the resistivity at zero field and higher magnetic field and the normal state Hall coefficient. The temperature dependent in plane resistivity at different doping has been studied [15]. The temperature and doping dependent resistivity data are consistent with published data [16]. The critical temperature $T_C$ has been determined from the temperature where the resistivity goes to zero at zero applied magnetic field. The resistive superconducting transition temperature $T_C$ shows the same dome-like behavior as published earlier [16]. LCCO shows very similar transport behavior as PCCO [17], but the superconducting dome is slightly shifted towards lower Ce concentration for LCCO. To understand the normal state behavior the resistivity has been measured for c-axis magnetic field $H>H_{C2}$ [15]. The resistivity shows a low T upturn for doping $x = 0.08$, $0.11$, and $0.13$, but, for $x>0.14$ no upturn is observed. The temperature dependent normal state Hall coefficient ($R_H$) of LCCO films measured at 14 T magnetic field shows that $R_H$ gradually changes from negative to positive value with increasing Ce concentration. $R_H$ is negative for the samples $0.08 \leq x \leq 0.13$ and positive for $x \geq 0.14$, which confirms that a Fermi surface reconstruction occurs at $x = 0.14$ [15].

## 2. *Vortex lattice melting in LCCO*

In many conventional and nonconventional superconductors a vortex liquid state is a region in the H-T plane which results from the melting of the vortex solid above some characteristic magnetic field, $H_{VS}$, as a consequence of thermodynamic fluctuation of the superconducting order parameter [18]. In cuprate superconductors the vortex liquid phase exists between a vortex solid phase below $H_{VS}$ (T) and the normal state above $H_{C2}$ (T) due to a strong 2D character and a low superfluid density [19].

We have measured the in-plane electrical resistivity versus magnetic field up to 14 T at different temperatures below $T_C$ for our LCCO films. $H_{VS}$ has been taken as the critical magnetic field (at different temperatures and doping) where the resistance becomes non-zero. As an example, Figure 1 shows a plot of the resistive transition versus field for $x = 0.15$ at different temperatures from 1.8 to 20 K. Previously it has been established in some hole-doped cuprates and the electron-doped PCCO [12] that there is no vortex liquid state at T = 0. Grissonnanche et al have shown that $H_{C2}= H_{VS}$ at T = 0 in the hole doped cuprates YBCO ($YBa_2Cu_3O_y$) and Y124 ($YBa_2Cu_4O_8$) from thermal conductivity measurements [20]. An expression for the vortex melting line due to thermal fluctuation has been derived using a nonlocal elasticity theory and Lindemann criteria by Houghton et al. for temperature dependence of $H_{VS}$ [21]:

$$\frac{\sqrt{b_m(t)}}{1-b_m(t)}\frac{t}{\sqrt{1-t}}\left[\frac{4(\sqrt{2}-1)}{\sqrt{1-b_m(t)}}+1\right]=\frac{2\pi c_L^2}{\sqrt{G_i}} \qquad (1)$$

where t =$T/T_C$, reduced field $b_m$=$H_{VS}(T)/H_{VS}(0)$, $G_i$ is the Ginzburg number, and $c_L$ is the Lindemann number. We have used the definitions of $G_i$ $c_L$ as used by Ramshaw et al. [22]. Figure 2 demonstrates the resulting field-temperature phase diagram for a typical Ce concentration, $x$ = 0.14. $H_{VS}$ (T) is extrapolated to T = 0 to get $H_{VS}$ (0) in the T = 0 limit. The black line is a fit to Eq. 1. The fitting of the data allows us to obtain the value of $H_{VS}$ (0) for different doping concentrations of Ce. We have plotted $H_{VS}$ (0) versus Ce doping in Fig. 11.

### 3. *Vortex Nernst effect in LCCO*

The vortex Nernst signal is defined as the transverse electric field generated by the vortices moving within the superconductor under the longitudinal thermal gradient and perpendicular magnetic field [23]. The vortex induced Nernst signal is well known from previous studies for conventional and high $T_C$ superconductors [1, 23, 24].

Figure 3 illustrates the typical magnetic field dependent Nernst signal, N=$E_y$/(-∇T) at different temperatures for the typical Ce doping (0.11) and (0.14) of LCCO thin films below $T_C$. The vortex induced Nernst signal peaks at a field H*. The position of H* shifts to the higher magnetic field with decreasing temperature, which follows the same behavior as the upper critical field $H_{C2}$ in the superconducting state below $T_C$. Above $H_{C2}$ the quasiparticle (normal state) Nernst signal is nearly linear in field but with a small additional $H^3$ term [17]. The normal state Nernst signal at higher field is subtracted from the measured data to obtain the net vortex signal as shown in the inset of Fig. 3 for $x$=0.14 at 20.7 K. We will discuss the quasiparticle Nernst signal later.

## 4. *Nernst effect in LCCO above $T_C$*

The nature of the Nernst signal observed just above the transition temperature $T_C$ is different from the Nernst signal well above $T_C$ i.e. in the normal state. Figure 4 (a) shows the raw Nernst signal as a function of field at T = 17.8 K (T> $T_C$=15 K) for $x$=0.15 (overdoped). The Nernst signal increases initially at low field with a subsequent fall at higher field. After a certain field the signal again is observed to increase with further increase in magnetic field. The signal at the higher magnetic field is attributed to the background signal from a sizable contribution of the normal state quasiparticles, which in general cannot assumed to be negligible.

Therefore the total Nernst signal is the sum of the Nernst signal, $N_{sc}$ due to superconducting fluctuations and $N_{qp}$ due to the quasiparticle contribution: $N=N_{sc}+N_{qp}$. We can estimate $N_{qp}$ from the total Nernst signal N at an applied magnetic field $H > H_{C2}$. To estimate $N_{qp}$ we have fitted the data above 10 T by using the power law: $N_{qp} = c_1$ (T) H+ $c_3$ (T) $H^3$. The same fitting procedure has been applied to the Nernst data for other underdoped, optimally doped and overdoped samples with slightly different values of $c_1$ and $c_3$ [17]. The dotted line is the fit of the high field segments of the above mentioned power law as shown in Fig. 4 (a). Thus we can extract the contribution due to superconducting fluctuation by $N_{sc} = N - N_{qp}$. Fig. 4 (b) shows the superconducting signal $N_{sc}$ as a function of magnetic field at the temperature of 17.8 K. The superconducting Nernst signal is observed to increase sharply at low magnetic field, reaches a maximum field H*, and then decreases gradually to a weakly temperature dependent magnitude. At high magnetic field $N_{sc}$ is vanishingly small.

Figure 5 presents the raw Nernst isotherm at different temperatures above $T_C$ for the typical concentration of Ce concentration, *x* = 0.14. For all the LCCO samples the superconducting Nernst signal shows the same behavior. Figures 6 shows the superconducting Nernst signal $N_{sc}$ as a function of field at different temperatures for underdoped (*x* = 0.08), optimal-doped (*x* = 0.11), and overdoped (*x* = 0.13, 0.14, and 0.16) samples. $N_{SC}$ is observed to decrease with increase in temperature above $T_C$ for all samples. There is no obvious change at the FSR doping. In contrast to the vortex peak in the superconducting state, the peak field H* shifts to the higher fields with increase in temperature, which indicates fundamentally distinct origins of Nernst signal above and below the transition temperature. H* is a characteristic field that was first identified by Kapitulnik et al. in disordered superconducting films of InGe [25]. Basically H*(T) (above $T_C$) mirrors the upper critical field $H_{C2}$ (T) (below $T_C$). For this reason these authors nicknamed it as the 'ghost critical field'. The ghost critical field is defined as the magnetic field scale above which the superconducting fluctuations are suppressed. Figure 7 shows H* (obtained from $N_{SC}$ versus H plot) as a function of reduced temperature ε for all the samples. H* obeys the logarithmic dependence

$$H^* = H_{C2}^* \ln (T/T_C) \tag{2}$$

where $H_{C2}^*$ is an empirical parameter that characterizes the strength of the superconductivity [5].

To analyze the superconducting fluctuation deep inside the normal state above the transition temperature the Nernst effect is a powerful tool. A large Nernst signal has been detected in a broad interval above $T_C$ in many hole-doped cuprates by the Princeton group in the period of 2000-2006 [1]. In the under-doped region the Nernst signal has been attributed to the superconducting phase fluctuation detectable up to ~ 5 $T_C$. They have defined the extended region having large signal above the "$T_C$ dome" as the Nernst region. The electron doped cuprate superconductors provide the interesting counterexample to the hole-doped cuprates. We have examined the Nernst region for electron doped LCCO films for the underdoped, optimally doped, and overdoped samples. The upper limit of the Nernst region is defined as $T_{onset}$. In other words $T_{onset}$ is defined as the temperature below which $\nu$ (T)/T starts to rise upon cooling. For example the data of LCCO for $x = 0.08$ yield $T_{onset} \cong 34.2$ K as shown in the Fig. 8. The phase diagram of $T_{onset}$ versus $x$ has been demonstrated in Fig. 9. $T_{onset}$ exactly follows the $T_C$ like dome. The phase diagram demonstrates the continuity of the region up to which the Nernst signal due to superconducting fluctuation is observed above $T_C$ dome for each doping.

## IV.  Discussion

In this section we will discuss the nature of the superconducting fluctuations in LCCO films. In addition the value of the upper critical field obtained from the Nernst data will be compared with the critical field obtained in the dirty limit from high-field transport properties. We also have determined the value of coherence length and the pairing gap in the LCCO thin films.

### 1. *Superconducting fluctuations*

The study of superconducting fluctuations had a remarkable revival after the discovery of the cuprate superconductors. Superconducting fluctuation can extensively be studied by the measurements of electrical resistivity, magnetization, and Nernst effect. The Nernst effect is a very sensitive probe to measure the superconducting fluctuation whereas it is very difficult to detect the superconducting fluctuation from the resistivity data. To understand the superconducting fluctuation and the strength of the superconductivity in electron doped LCCO samples we have analyzed the data obtained from Nernst measurements.

To treat the fluctuation of the superconducting order parameter in the Gaussian approximation, the off-diagonal Peltier coefficient $\alpha_{xy}$ is an important thermodynamic state function [11]. In the zero field limit $\alpha_{xy}/H$ is simply related to the Nernst coefficient $\nu$ and the electrical conductivity $\sigma_{xx}$ of the sample through the formula $\alpha_{xy}/H \approx \nu \sigma_{xx}$, when the Hall conductivity is small. Normally in cuprates the transverse Hall conductivity is very small [26, 12]. In superconductors, above the transition temperature the conductivity varies weakly with the change of the temperature and magnetic field. So the Peltier coefficient is mostly governed by the magnitude of the Nernst coefficient. We have plotted the Nernst coefficient $\nu$ ($\nu = N/H$) versus magnetic field for the optimally doped sample ($x=0.11$) at different temperatures above $T_C$ as shown in Fig. 10 (a). We can see that $\nu$ is independent of magnetic field and is nearly constant in low magnetic field limit. From $\nu$ ($\nu = N/H$) versus H plots for all the samples the value of $\nu$ ($\nu_0$) in the zero magnetic field limit has been extracted and plotted as the function of reduced temperature $\varepsilon$. The plot of $\nu_0$ as a function of $\varepsilon$ is shown in the Fig. 10 (b) for $x=0.11$. To estimate the normal state contribution, $\nu_0^{qp}$ in the limit $H\rightarrow 0$ we have fitted the data and extracted the superconducting contribution, $\nu_0^{sc} = \nu_0 - \nu_0^{qp}$ by using the power law: $\nu_0^{qp} = A(T) T + B(T) T^2$, where A and B are constants [5]. The same fitting procedure has been applied to the Nernst data for other underdoped, optimallydoped and overdoped samples with slightly different values of A and B. The solid black line as demonstrated in Fig. 10 (b) represents the quasiparticle background to the Nernst coefficient. In Fig. 10 (c) we have displayed the superconducting contribution to $\nu_0$ called $\nu_0^{sc}$ versus reduced temperature for $x=0.11$ after subtracting the quasiparticle contribution $\nu_0^{qp}$.

*Comparison to temperature dependent Gaussian theory*

We have calculated the contribution of Gaussian superconducting fluctuations to the transverse thermoelectric response above $T_c$ in the low magnetic field limit. However it should be emphasized that the temperature up to which the fluctuation can be detected is a question of sensitivity and signal-to-background ratio. In this study we have tracked the fluctuations up to ~ 1.6 $T_C$ for LCCO thin films. The value of $\alpha_{xy}^{SC}$ has been determined experimentally using the equation:

$$\frac{\alpha_{xy}^{SC}}{H} \approx \frac{\nu_0^{SC}}{\rho_s} \quad (3)$$

where the superconducting Nernst coefficient $\nu_0^{SC} = \nu_0 - \nu_0^{qp}$ in the zero field limit and $\rho_s = \rho/s$, s is the interlayer separation (6.1 Å) . Figure 10 (c) displays $\nu_0^{SC}$ as a function of reduced temperature ε for *x*=0.11.

Superconductivity is normally characterized by the superconducting order parameter SOP (|Ψ| $e^{i\theta}$) which comprises the amplitude Ψ and the phase θ at each space point [27]. Fluctuation in either amplitude or phase or both amplitude and phase can affect the superconducting properties and Nernst signal can be enhanced [28]. Superconducting fluctuations deal with the thermal fluctuations of the SOP [29]. By using the time dependent Ginzburg-Landau equation in the Gaussian approximation, Ussushkin et al. calculated the off-diagonal component of the Peltier conductivity tensor, $α_{xy}$, which results from the contribution of Gaussian superconducting fluctuations to thermoelectricity [2]. The value of $α_{xy}^{SC}$/H is independent of magnetic field at low magnetic field. In 2009, Serbyn et al. calculated $\alpha_{xy}^{SC}$ for 2D type II dirty *s*-wave superconductor independently up to arbitrary T and arbitrary H [30, 11]. In 2012 and 2014, calculation of $\alpha_{xy}^{SC}$ has been performed for hole-doped Eu-LSCO and electron doped PCCO, respectively [5, 12].

We have compared the temperature dependence of $\alpha_{xy}^{SC}$ measured for our LCCO thin films with the predictions of Gaussian theory in the low field limit (H→0). In the low field limit, the Aslamazov-Larkin theory of superconducting fluctuation predicts that

$$\frac{\alpha_{SC}^{xy}}{H} = \nu_{SC}\sigma_{xx} \sim \frac{\xi_0^2}{T\ln(T/T_C)} \quad (4)$$

where $\xi_0$ is the Ginzburg-Landau coherence length at zero temperature. Therefore for a given T, $\alpha_{SC}^{xy}$ depends only on one quantity $\xi_0$. The extracted values of $\frac{\alpha_{SC}^{xy}}{H}$ at zero magnetic field are shown in Fig. 12 for $x = 0.08$, 0.11, and 0.14. As shown in Fig. 12 the data are observed to follow the theoretical temperature dependence for underdoped, optimal doped and overdoped samples. The agreement between theoretical and experimental data is excellent. It is evident from Fig. 12 that the Gaussian theory can reliably explain the superconducting fluctuation from the underdoped to the overdoped regime.

## 2. *$H_{C2}$ (0) and $H_{C2}$\**

The doping dependence of $H_{C2}$ has been reported for hole-doped and electron-doped cuprates using high-field measurements of $H_{VS}$ (T). For Tl-2201 ($Tl_2Ba_2CuO_{6+\delta}$) and YBCO it has been shown that the $H_{VS}$ (0) extrapolated from the resistive $H_{VS}$ (T) is consistent with the $H_{C2}$ (0) value obtained from a low temperature thermal conductivity measurement [20]. The field temperature phase diagram for both YBCO and Y124 ($YBa_2Cu_4O_8$) demonstrates that $H_{C2}$ (T) and $H_{VS}$ (T) converge at T=0 K, where $H_{VS}$ (T) is obtained from high field resistivity data and $H_{C2}$ (T) has been determined from the thermal conductivity measurement. So from that measurements of $H_{VS}$ vs T one can determine $H_{C2}$ (0). The data $H_{C2}$ (0) = $H_{VS}$ (0) has also been confirmed for the electron doped PCCO at $x$=0.15 [12]. Based on these prior investigations we take $H_{VS}$ (0) as $H_{C2}$ (0) in the present study.

We have compared the critical field in the LCCO films in two different ways: a) from high field transport measurements below $T_C$, denoted as $H_{C2}$ (0) as discussed above and b) $H_{C2}$\* directly obtained from the Nernst data above $T_C$. The two measured values of critical field are in reasonable agreement. Both have their highest value near optimal doping. With increasing doping the magnitude is observed to decrease and in case of underdoped sample the value of upper critical field is also lower than the optimal doped one as shown in Fig. 11.

The variation of $H_{C2}^*$ has been displayed in Fig. 11. To determine $H_{C2}^*$, the ghost critical field $H^*$ has been plotted versus reduced temperature $\varepsilon = (T-T_C)/T_C$ for the samples as shown in Fig. 10. $H^*$ obeys Eq. 2 for all the samples up to high temperature namely the reduced temperature of $\varepsilon \approx 0.6$. Below $\varepsilon \approx 0.15$ $H^*$ deviates from the relation of Eq. 2. This deviation has been attributed to the divergence of paraconductivity in $\varepsilon \rightarrow 0$ limit [2]. The Nernst signal $N_{SC}$ is the ratio of the off diagonal Peltier coefficient ($\alpha_{xy}$) from superconductivity fluctuation to the electrical conductivity ($\sigma_{xx}$). In the $\varepsilon \rightarrow 0$ limit both $\alpha_{xy}$ and $\sigma_{xx}$ diverge [2]. Using Eq. 2 the values of $H_{C2}^*$ have been determined for the underdoped, optimal doped, and overdoped samples. These values can be compared with the estimated upper critical field $H_{C2}(0)$, which has been obtained as the resistive critical field $H_{VS}(0)$ at shown in Fig. 2. The value of $H_{C2}^*$ extracted from the fit for each doping has been plotted in Fig. 11. The main finding is that the field scale of the superconductivity decreases with decreasing doping with a local maximum at optimal doping and again decreases with increase above optimal doping. The obtained result is free from any theory or model and can be directly read off from the raw isothermal Nernst data as a function of magnetic field. $H_{VS}(0)$ has been observed to be higher than upper critical field $H_{C2}^*$ for optimally ($x=0.15$) doped $Pr_{2-x}Ce_xCuO_4$ (PCCO) [12]. The anomaly in $H_{VS}(0)$ seen in PCCO has been linked to the Fermi Surface reconstruction (FSR) ($x=0.16$). Similar trend of $H_{VS}(0)$ is observed near optimal doped LCCO ($x=0.11$) as reported in our manuscript. However the Fermi surface reconstruction in LCCO is at $x=0.14$, which is much higher than the optimal doped LCCO where $H_{VS}(0)$ follows $H_{C2}^*$. From our finding it is not possible to link the FSR to cause of $H_{VS}(0)$ anomaly. It is not clear that why the values are different near optimally electron doped superconductor. However $H_{VS}(0)$ and $H_{C2}^*$ closely track the superconducting dome.

### 3. *Determination of $\xi_0$ and $\Delta_0$*

We carried out the scaling comparison for the over-doped, optimal-doped and under-doped samples of LCCO films. In type II superconductors the upper critical field $H_{C2}$ is an important parameter which determines the pairing gap $\Delta_0$ through the coherence length $\xi_0$ i.e. the size of the Cooper pair and the strength of the Cooper pairing potential [9]. The pairing potential is stronger and the pair size is smaller for higher $H_{C2}$. Below $H_{C2}$ the vortex liquid state appears in the sample. But the vortex liquid immediately forms a lattice and thus the electrical resistance is going to be zero and the vortex melting field $H_{VS}$ is equal to $H_{C2}$. In cuprate superconductors the vortex liquid phase intervenes between the vortex solid phase and the normal state. The value of the pairing gap is a polarizing issue for understanding the strength of the superconductivity in cuprates. So far, different evolution of pairing gap and the estimation of upper critical field $H_{C2}$ are in sharp contradiction [5, 31].

According to the BCS theory the energy gap to the onset of single particle excitations decreases from a certain value at T=0 K to zero at T=$T_C$ [32]. In the case of a d-wave superconductor the gap according to BCS is, $\Delta_0 = 2.14 k_B T_C$ [33]. The behavior of the gap amplitude for all doping of LCCO is shown in Fig. 13. Just like $Pr_{2-x}Ce_xCuO_{4-\delta}$ the superconducting gap follows the doping dependence of $T_C$ [34]. The behavior of $H_{C2}*$ in our experiment is consistent with the behavior of the gap amplitude at different Ce doping as shown in Fig. 11 and 13.

Now we will discuss about the determination of the coherence length $\xi_0$ for different dopings directly using the value of $H_{C2}*$. In the case of cuprate superconductors the strongly correlated background is weakened by increased doping. The behavior of various superconductivity characteristics, superconducting fluctuation and their predominance in the phase diagram depend on different material dependent parameters like quenched disorder, dimensionality or the superconducting coherence length etc. [35]. The upper critical field is closely related to the superconducting gap magnitude $\Delta_0$. The upper critical field and the superconducting gap can be compared directly by converting them to the length scales. The gap amplitude can be converted to the Pippard coherence length by

$$\xi_p = \hbar v_F / a \Delta_0 \quad (5)$$

where $v_F$ (2.9 × 10$^5$ m/s, for the overdoped n-type cuprate NCCO [36] is the Fermi velocity, $a$ = 1.5 (for $d$-wave superconductor) and $\Delta_0$ is the superconducting gap amplitude. The value of Fermi velocity has been determined from the Quantum oscillation measurements. $\Delta_0$ is 2.14 $k_B T_C$ for a d-wave state. In a dirty superconductor the Pippard's relation for the coherence length is

$$1/\xi_0^{dirty} = 1/\xi_p + 1/l \quad (6)$$

where $l$ ($l = hs/e^2 \rho_0 k_F$) is the mean free path, $s$ (6.1 Å) is interlayer distance, $k_F$ (0.58 Å$^{-1}$) is the Fermi wave vector, and $\xi_0^{dirty}$ is the coherence length at T=0. $\rho_0$ = 17 μΩ cm is residual resistivity for $x$=0.16. We get the value of $\xi_0$ as 12.5 nm which is consistent with the previous results in the overdoped PCCO, $x$=0.17 ($\xi_0$ was calculated to be 10.9 nm) [13]. We can estimate the value of upper critical field by using the value of the coherence length in the dirty limit

$$H_{C2} = \varphi_0 / 2\pi (\xi_0^{dirty})^2 \quad (7)$$

$H_{C2}$ is calculated to be 2.1 T for 0.16 sample. For overdoped PCCO, $H_{C2}$ was observed to be 3.7 T [13]. We get the upper critical field of 2.5 T from Nernst measurement and $H_{VS}$ (0) of 2.8 T from high field magnetoresistance measurement for $x$ = 0.16. So we can see that the value of upper critical field obtained from fluctuation above $T_C$ is in good agreement with the critical field expected for the sample in the dirty limit and the high-field transport directly measured at $T < T_C$.

According to Pourret et al. at a temperature above $T_C$ for the conventional superconductor $Nb_{0.15}Si_{0.85}$ there is a magnetic field called the ghost critical field below which the superconducting fluctuation is controlled by the coherence length $\xi(T) = \xi(0)/(\ln(T/T_C))^{1/2}$ and above which their extent is controlled by the magnetic length $l_B = (\hbar/2eB)^{1/2}$ [9]. At the ghost critical field the two length scales become the same, i.e. $\xi(H^*) = l_B(H^*)$, i.e, $H^* = (\varphi_0/2\pi\xi_0^2)\ln(T/T_C)$, where the flux quantum $\varphi_0 = h/2e$ and $H_{C2}(0) = (\varphi_0/2\pi\xi_0^2)$. As we have discussed, $H_{C2}(0)$ is nearly equal to $H_{C2}^*$, so we can roughly estimate the value of the coherence length $\xi_0$ from $H_{C2}^* = (\varphi_0/2\pi\xi_0^2)$. We thus can determine the value of the coherence length $\xi_0$ for different doping directly using the value of $H_{C2}^*$. The inset of Fig. 13 displays the coherence length at different doping concentrations. From the calculated value of $H_{C2}^*$ and $\xi_0$ we can say that the strength of the superconductivity develops and the coherence length vanishes simultaneously.

The present experiment establishes that the upper critical field increases as Ce doping decreases, is highest for the optimal doped sample and again decreases for underdoped sample, which implies that the superconductivity is strongest for the optimally doped sample. The trend of $\Delta_0$ suggests the dome shape of the entire cuprate phase diagram.

## 4. *Possible origin of the $T_C$ dome*

We attribute the weakening of the strength of the superconductivity and the fall of the transition temperature $T_C$ with underdoping as coming from competition with the onset of antiferromagnetic order. Very few experiments have been performed in $La_{2-x}Ce_xCuO_4$ because it can only be made in a thin-film form and no crystals can be prepared. So neutron scattering, NMR and bulk μSR are not possible for a study of the magnetism of LCCO. Recently from a depth resolved low energy μSR study of $La_{2-x}Ce_xCuO_4$ films the presence of long range AFM order has been observed up to just below optimal doping [37]. The static 3D magnetism (e.g.

long range antiferromagnetism) is absent in the $x = 0.10$ sample, but present in $x = 0.07$ below 65 K and $x = 0.08$ below 40 K. Above $x = 0.10$ there is short range AFM order. The transition of long range antiferromagnetic order to short range antiferromagnetic order is also observed for NCCO from inelastic neutron scattering measurements [38]. Thus the competition with the long range AFM order weakens the superconductivity and both $H_{C2}$ and $T_C$ fall in the underdoped $La_{2-x}Ce_xCuO_4$.

## 5. $T_{onset}$

In hole-doped cuprates the Nernst results suggested that above $T_C$ superconductivity disappears because of the loss of long range phase coherence [1]. At a temperature called $T_{onset}$ the Nernst region in these cuprates does not extend to the pseudogap temperature T* but lies between $T_C$ and T* for underdoped samples. For underdoped samples $T_{onset}$ falls steeply when $x$ tends to zero, whereas T* is observed to increase continuously. Below $T_{onset}$ the presence of the Nernst signal and the diamagnetic response are distinct signatures of fluctuating superconductivity [1]. For electron doped cuprates a pseudogap phase has not been observed, only a region of AFM fluctuations above $T_C$ is observed. In the superconducting dome all the LCCO films exhibit a fluctuation Nernst signal above $T_C$ up to a characteristic crossover temperature $T_{onset}$ as shown in Fig. 9. For LCCO thin films the optimally doped ($x=0.11$) sample has the highest temperature range and thus $T_{onset}$ ($\cong$ 39 K) is high for $x=0.11$. For higher doping $T_{onset}$ decreases and tends to zero along with the critical temperature at the end of the superconducting dome. Both $T_{onset}$ and $H_{C2}$ are observed to track the $T_C$ dome in LCCO. It seems that, over the broad interval $T_{onset}>T>T_C$, the superconducting pairing strength and the superconducting fluctuation may be intimately related. Similarly in hole-doped LSCO the Nernst region has been observed to be diminished at the end of the superconducting dome [1] implying that there is a common relation between superconducting fluctuation and pairing in both electron and hole-doped cuprates.

## V. Summary and Conclusion

We have performed measurements of the Nernst effect in electron-doped $La_{2-x}Ce_xCuO_4$ films over a wide range of doping and temperature above and below $T_C$. Above the superconducting transition temperature the field dependent Nernst signal reveals a field scale that increases with increasing temperature. We have uncovered an extended region above the superconducting dome where the Nernst signal exists. The upper limit of the Nernst signal has been defined by $T_{onset}$ which is high for optimally doped LCCO film. The Nernst signal above the critical temperature displays a smooth continuity with the vortex liquid melting field at zero temperature.

By subtracting the quasiparticle contributions to zero in the normal state of the LCCO films we have resolved a sizable Nernst signal coming from fluctuating Cooper pairs. We also have elucidated the nature of superconducting fluctuation from the Nernst measurements above critical temperature. A quantitative agreement with the theoretical prediction of Gaussian fluctuations in the zero field limit has been observed for all doping.

We have extracted the characteristic field $H_{C2}^*$ directly from the measurements of the Nernst effect, which determines the value of coherence length i.e. the size of the Cooper pair and the strength of pairing potential. The pairing potential is stronger and the pair size is smaller for optimal doped ($x=0.11$) LCCO thin films. A dome like doping dependence of pairing potential, similar to the $T_C$ dependence, is observed. The weakening of the pairing strength and the fall of the transition temperature $T_C$ with underdoping has been attributed to the onset of antiferromagnetic order. Surprisingly, no special feature in the pairing potential occurs at the doping ($x=0.14$) where the Fermi surface undergoes a reconstruction for LCCO films.

**Acknowledgments**: This research was supported by the NSF under DMR-1708334, the Maryland "Center for Nanophysics and Advanced Materials". The National Institute of Standards and Technology, U.S. Department of Commerce, in supported this research through Awards No. 70NANB12H238 and No. 70NANB15H261.

**Figures**

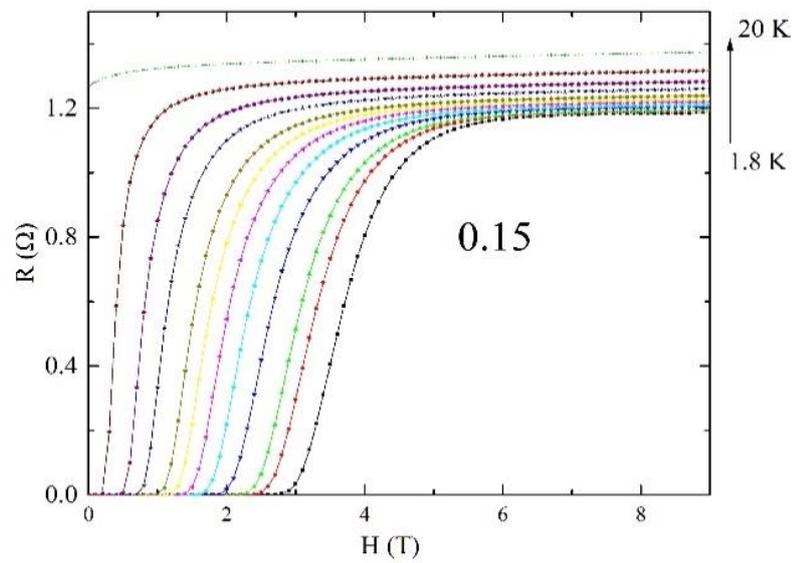

**Figure 1**. The resistance of LCCO films for $x = 0.15$ as a function of magnetic field, from 1.8 to 20 K. The onset of resistivity as the magnetic field is increased, marks the vortex lattice melting transition, $H_{VS}$ (T).

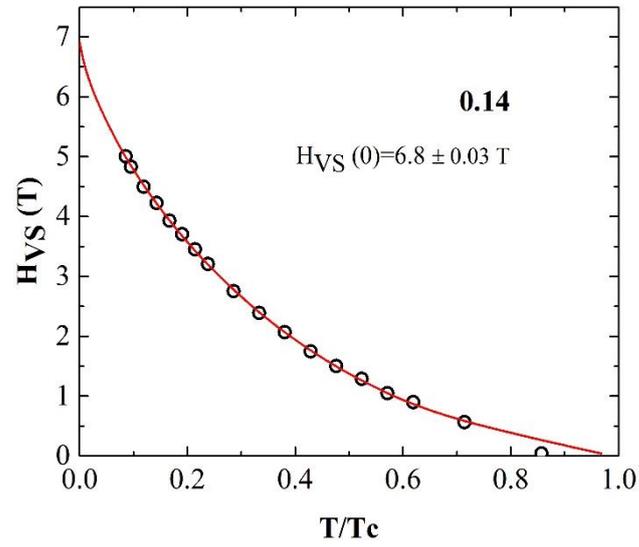

**Figure 2**. The vortex solid melting transition as a function of temperature for $x$ = 0.14. The temperature axis is scaled by $T_C$. The solid lines are the best-fit lines to Eq. 1 i.e. in text $H_{VS}$ (0) is obtained from the extrapolation of the best-fit line of $H_{VS}$ (T) to T = 0.

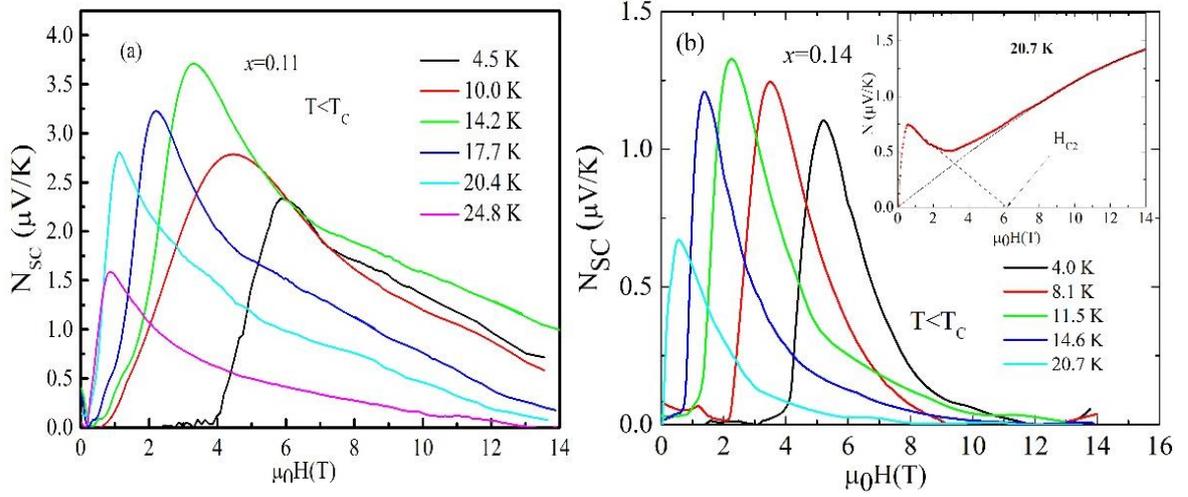

**Figure 3**. (a) The superconducting Nernst signal $N_{SC}$ of LCCO as a function of magnetic field below $T_C$ for $x$= 0.11 and 0.14 with $T_C$ =0.26 and 21 K, respectively as determined by the onset of the resistive transition. The peak field, where the maximum of $N_{SC}$ is obtained with quasiparticle background subtracted. Inset shows a typical raw Nernst signal N versus magnetic field H. The black dotted line represents the polynomial fit using the power law $N_{qp} = c_1 (T) H + c_3 (T) H^3$, where $N_{qp}$ is the quasiparticle background. (b) $N_{SC}$ of LCCO as a function of magnetic field below $T_C$ for $x$= 0.11 with $T_C$ =26 K.

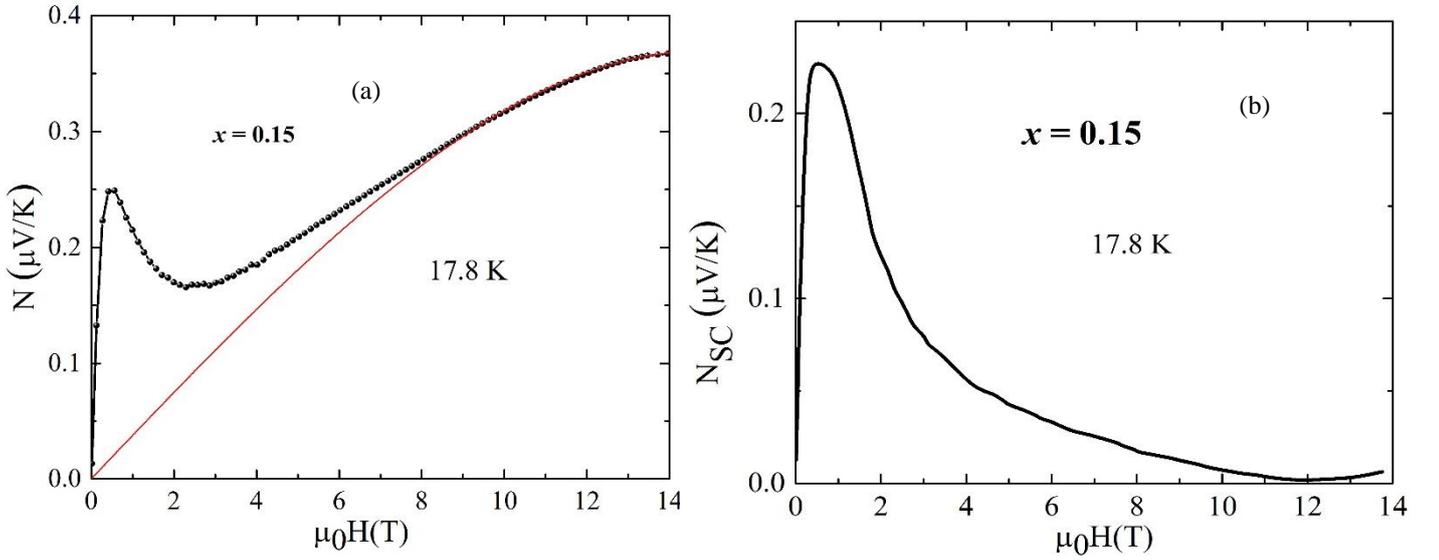

**Figure 4**. (a) Raw Nernst signal N as a function of field H for $x = 0.15$ at T=17.8 K. The red line is the polynomial fit to the raw data above 10 T, of the form $N_{qp} = c_1(T) H + c_3(T) H^3$. (b) Superconducting contribution to the Nernst signal as obtained by subtracting the normal state quasiparticle Nernst signal $N_{qp}$ from the raw Nernst response for $x = 0.15$. H* represents the peak field in $N_{SC}$ versus H plot above $T_C$.

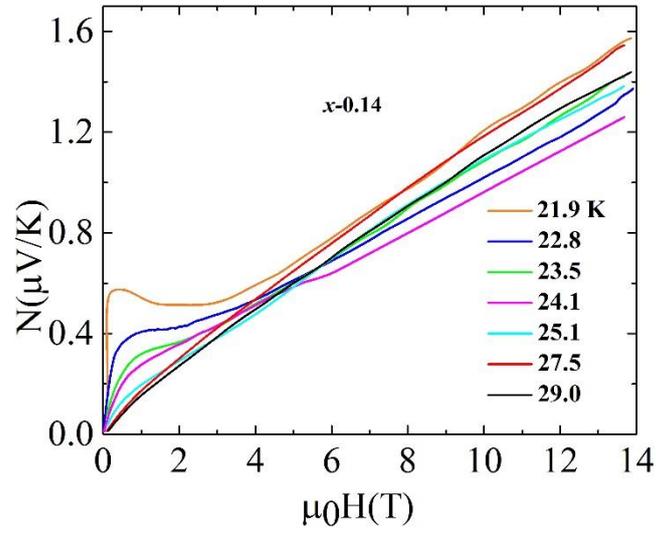

**Figure 5**. Raw Nernst data as a function of field for *x* = 0.14 at different temperatures above T$_C$.

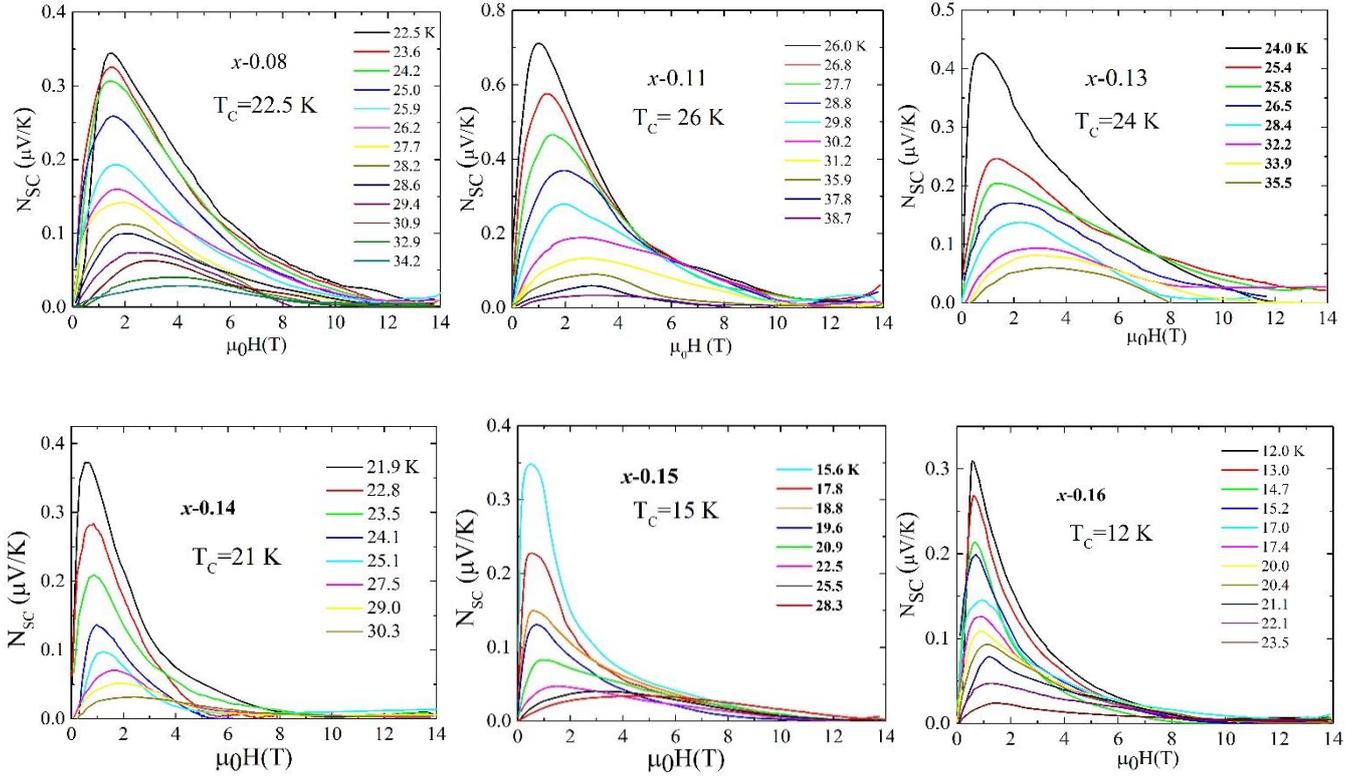

**Figure 6**. Superconducting contribution to the Nernst signal at different temperatures above $T_C$ for all doping. The peak field shifts towards higher field with increasing temperature.

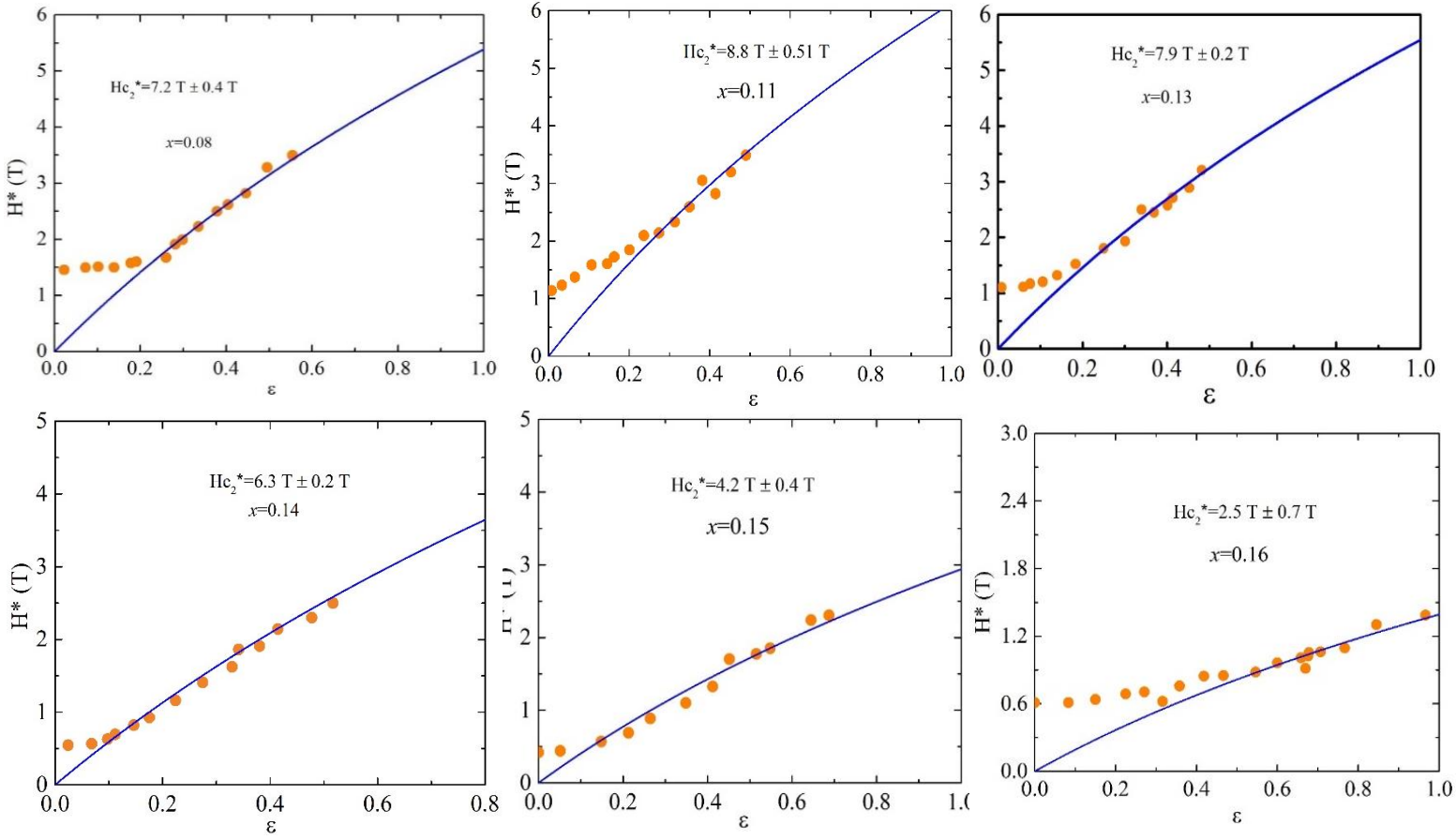

**Figure 7**. The field scale H* as a function of reduced temperature ε = (T-T$_C$)/T$_C$ in LCCO at doping as indicated. The solid lines are the fit to the function of logarithmic dependence H*=H$_{C2}$* ln (T/T$_C$). We can extract the characteristic field H$_{C2}$* from the fittings for the samples.

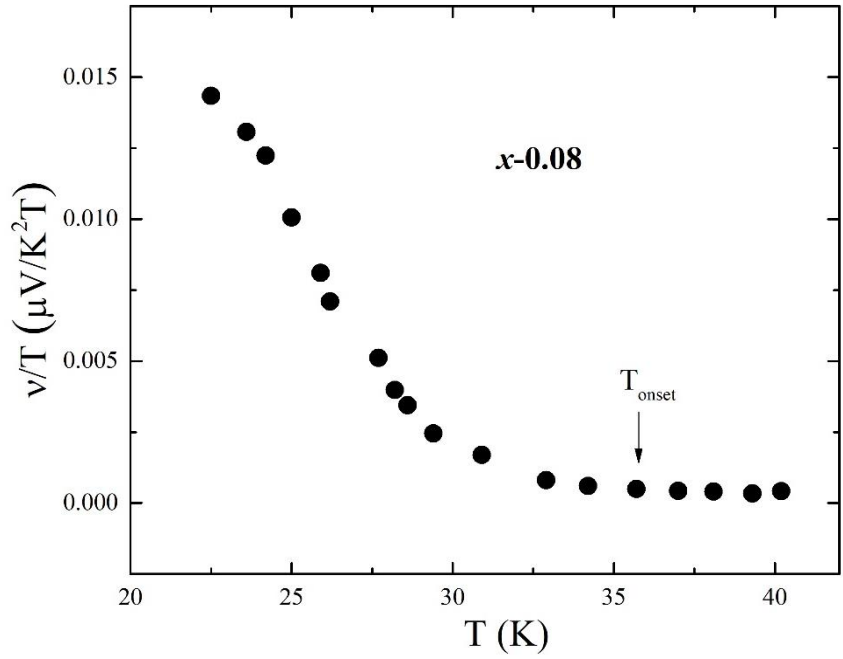

**Figure 8**. Nernst coefficient ν of LCCO at the electron doping *x* = 0.08, plotted as ν/T versus temperature T at a magnetic field of 2 T. The onset temperature $T_{onset}$ is indicated by the arrow.

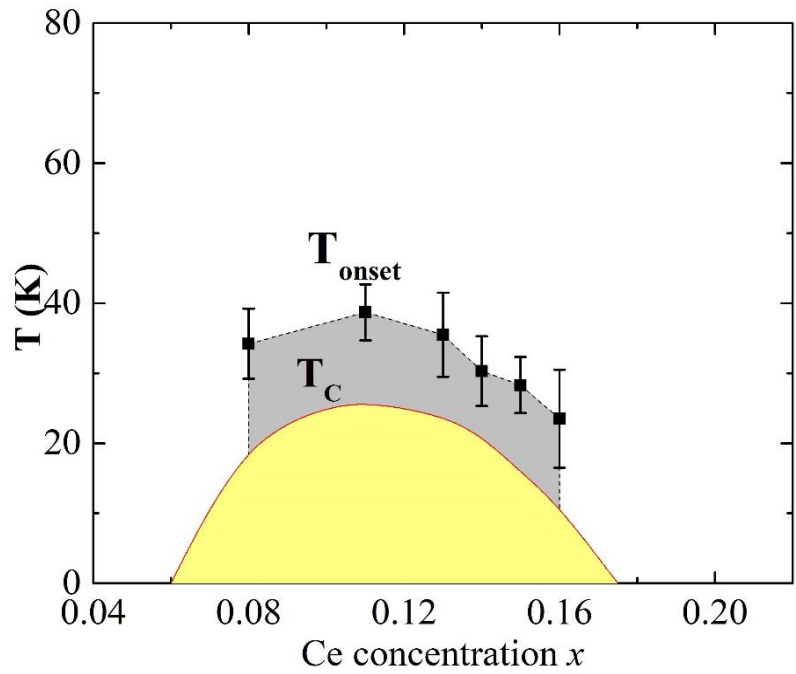

**Figure 9**. The phase diagram of LCCO showing the SC fluctuation Nernst region between the $T_C$ dome and $T_{onset}$ based on Nernst measurements for six different doping. $T_{onset}$ tracks the $T_C$ dome.

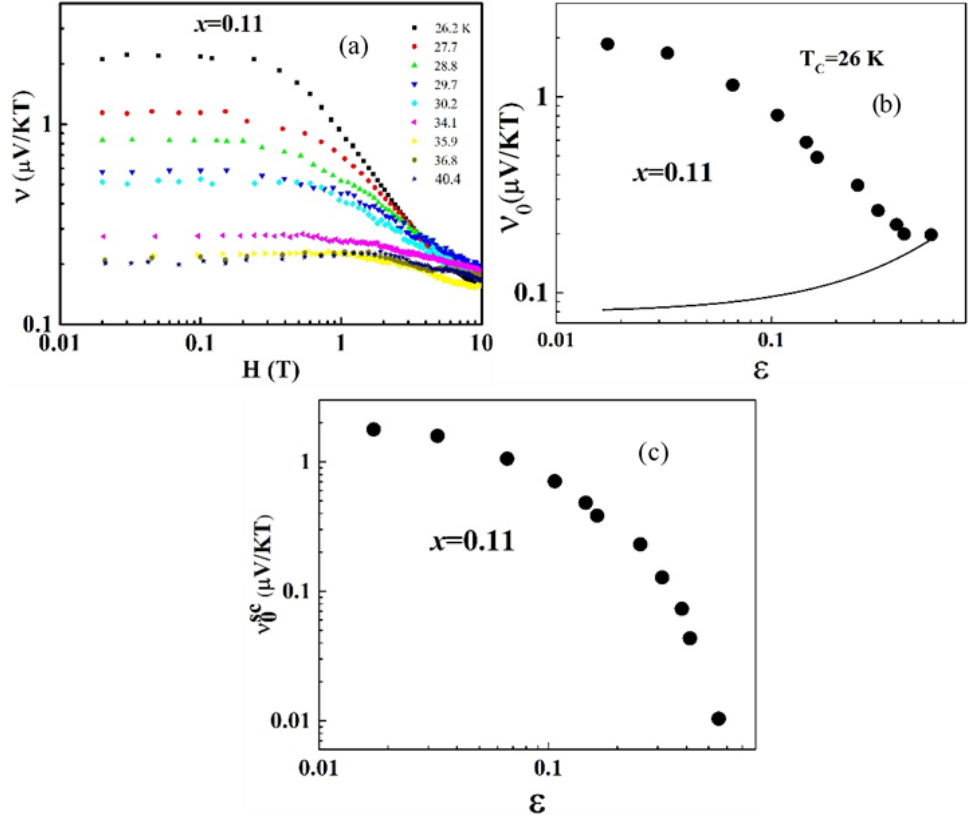

**Figure 10**. (a) Nernst coefficient as a function of Magnetic field for the optimal doped sample (Ce=0.11) measured at temperature exceeding $T_C$. As $H \to 0$, the Nernst coefficient is independent of magnetic field with a constant value $\nu_0$. At higher magnetic field the Nernst coefficient becomes independent of temperature. (b) $\nu_0$ as a function of reduced temperature $\varepsilon$ for $x = 0.11$. The solid line is the normal state contribution ($\nu_{qp}$) to $\nu_0$. (c) $\nu_0^{SC}$ (obtained by subtracting the $\nu_{qp}$ from the Nernst coefficient $\nu_0$) as a function of reduced temperature $\varepsilon$ for $x = 0.11$.

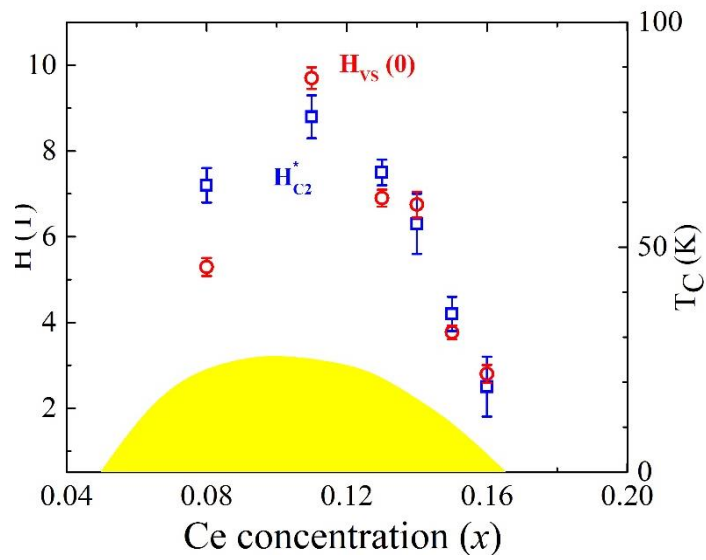

**Figure 11**. The phase diagram of two magnetic field scales of the superconductivity in LCCO plotted as a function of Ce doping.

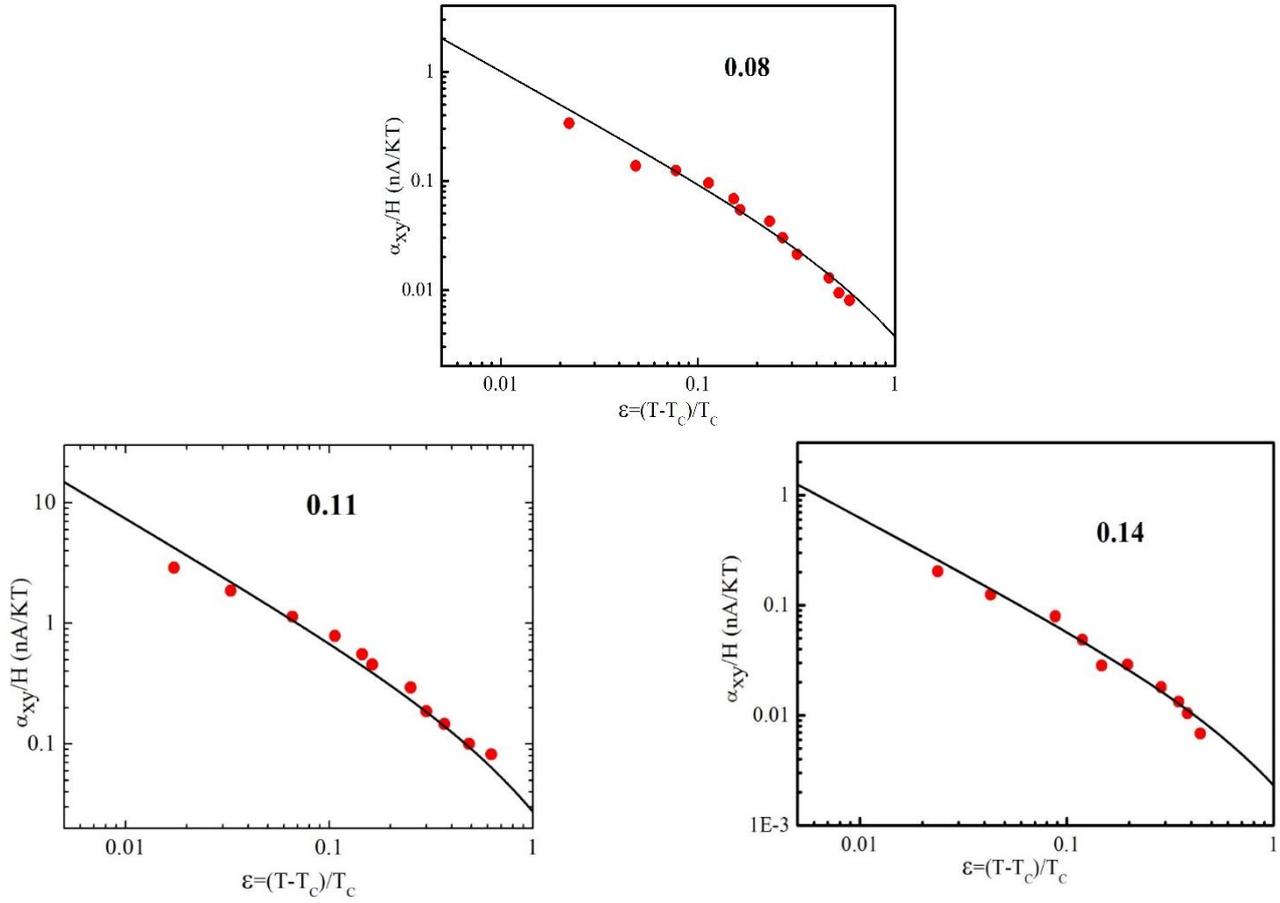

**Figure 12**. $\alpha_{xy}/H = \nu/\rho_{xx}$ in the zero field limit extracted from the measured Nernst coefficient and resistivity in LCCO for $x = 0.08, 0.11$, and $0.14$. The solid line is the theoretical expectation for the zero field limit from Gaussian fluctuations.

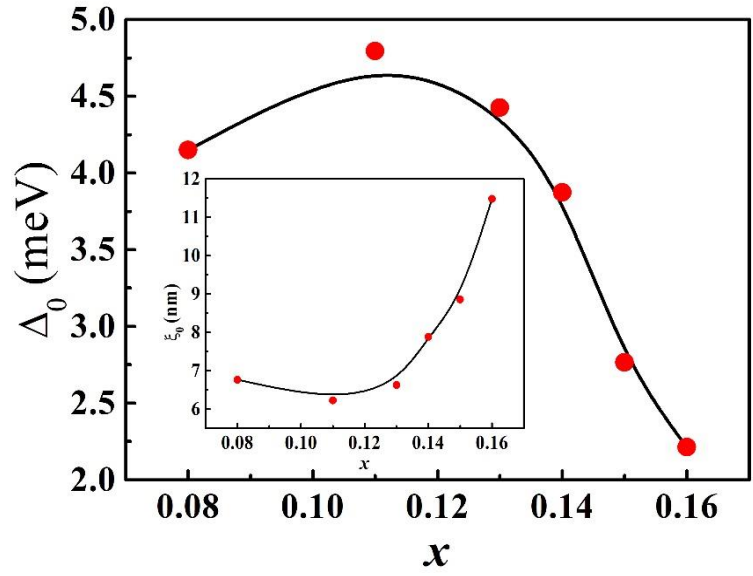

**Figure 13**. Ce dependence of superconducting gap. Inset shows the coherence length obtained from $H_{C2}^*$.